# Comment on "Bounding and approximating parabolas for the spectrum of Heisenberg spin systems" by Schmidt, Schnack and Luban

O. Waldmann
Physikalisches Institut III, Universität Erlangen-Nürnberg, D-91058 Erlangen, Germany

Recently, Schmidt et al. proved that the energy spectrum of a Heisenberg spin system (HSS) is bounded by two parabolas, i.e. lines which depend on the total spin quantum number S as S(S+1) [1]. The prove holds for homonuclear HSSs which fulfill a weak homogenity condition. Moreover, the extremal values of the exact spectrum of various HSS which were studied numerically [1,2] were found to lie on approximate parabolas, named rotational bands, which could be obtained by a shift of the boundary parabolas. In view of this, it has been claimed that the rotational band structure (RBS) of the energy spectrum is a general behavior of HSSs. Furthermore, since the approximate parabolas are very close to the true boundaries of the spectrum for the examples discussed, it has been claimed that the methods of ref [1] allow to predict the detailed shape of the spectrum and related properties for a general HSS.

In this comment I will show by means of examples that the RBS hypothesis is not valid for general HSSs. In particular, weak homogenity is neither a necessary nor a sufficient condition for a HSS to exhibit a spectrum with RBS.

First, I will discuss two HSS which fulfill the weak homogenity condition. Figure 1a shows the spectrum of a hexanuclear ring "doped" with a central spin. The ratio of interring coupling J and ring-to-central-spin coupling $J_1$ is set to $\alpha \equiv J_1/J = 0.4$ in order to ensure weak homogenity. At first sight, the approximate parabolas of ref [1] seem to be acceptable. However, for J > 0 the ground state belongs to S = 1. The first excited state belongs to S = 0 with an excitation energy of 0.194 J while the lowest S = 2 state has an excitation energy of 0.503 J. Thus, for this system the approximate parabolas cannot correctly describe the low-temperature properties, i.e. here the RBS hypothesis does not allow for a detailed prediction of the properties of the HSS. This is more evident for the HSS indicated in Fig. 1b (and for all the other examples below). $\alpha = 0.2$ ensures weak homogenity. While for the HSS of Fig. 1a the approximate parabolas appear as roughly correct, here both the upper and lower approximate parabolas are clearly unacceptable representations of the true extremal energies, in particular for states with S considerably smaller than $S_{max} = \Sigma s_\mu$. On the other hand, the spectrum of HSSs which are not weakly homogenous nevertheless may be well approximated by the RBS hypothesis. For instance, Fig. 1c shows the spectrum of a hexanuclear ring where the coupling strength of one bond has been increased from J to 2J.

One deficiency of the "boundary line" type of argumentation used in ref [1] is illustrated by the tetranuclear star (Fig. 1d). Here, the energies can be calculated exactly as $E(S, S_{234}) = J/2\, S(S+1) - E(S_{234})$. Whereby, $E(S_{234}) = J/2\, S_{234}(S_{234}+1)$, $S_{234} = S_2 + S_3 + S_4$, $S = S_1 + S_{234}$. Although the energies strictly fulfill $E(S) \propto S(S+1)$, the true boundaries are only partly described by the approximate parabolas since the allowed values of S are restricted by spin coupling rules. Obviously, this is only the most trivial effect not accounted for by the reasoning of ref [1]. There must be further effects which disturb the true boundaries from a simple S(S+1) behavior, as is evident from e.g. the spectrum of the HSS shown in Fig. 1e.

It has been noted in ref [1] that the approximate parabolas exactly meet the extremal energies for $S = S_{max} - 1$. This is not accidental. Actually, the prove can be found in most text books dealing with spin-waves, e.g. [3]. Starting with the states $\sqrt{2s_\mu}|\mu\rangle = S_\mu^-|F\rangle$ as basis for the subspace with magnetic quantum number $M = S_{max} - 1$, the calculation of the energies of the ferromagnetic and spin-wave states [4] reduces to a diagonalization of the matrix $J_{\mu\nu}s_\mu s_\nu$ in the general case - or, respectively, of $J_{\mu\nu}$ in the case of homonuclear HSS as found in ref [1]. This result demonstrates that the approximate parabolas of ref [1] are nothing else than extrapolations of the energies of the ferromagnetic and the spin-wave states to smaller values of S. In particular, they do not take into account the effects of quantum fluctuations, i.e. those effects which make HSS nontrivial. It should be noted that nowhere in this discussion the condition of weak homogeneity is required.

Obviously, the RBS hypothesis in the sense of ref [1] is not a general feature of HSSs. To my opinion, the significance of the notion of rotational bands comes from the observation that in several cases *parts* of the spectrum can be reasonably approximated by rotational bands, or a set of rotational bands, respectively [5]. If this is in particular true for the lowest lying energy levels then also the low-temperature properties can be predicted. However, the curvature of these bands is typically strongly affected by quantum fluctuations, as has been discussed, e.g., in detail for Heisenberg rings [5]. The curvatures given in ref [1] are useful in general only for states close to the ferromagnetic state, i.e. if S is close to $S_{max}$. For states with small S, the curvature even may have the opposite sign than the predictions of ref [1] (inspect Figs. 1b, 1e). Certainly, there are HSS for which the approximate parabolas of ref [1] are good approximations, but ref [1] does not provide a criterion to pick them out (despite the obvious criterion $s \to \infty$).

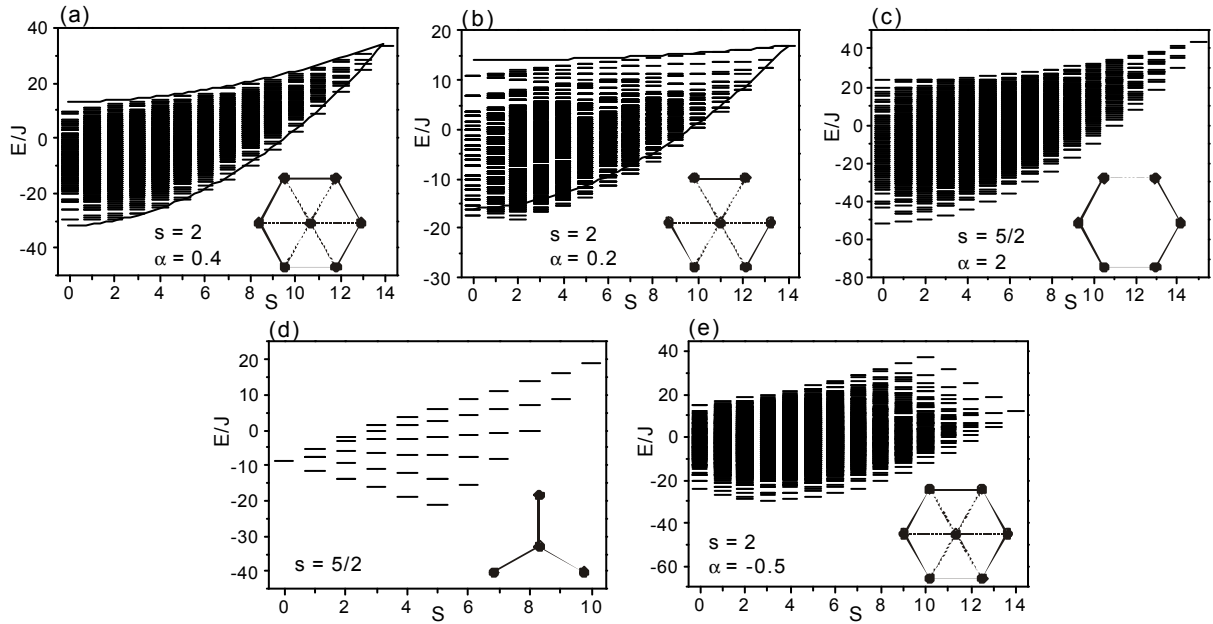

Fig. 1 – Energy spectrum of the Heisenberg spin systems discussed in the text. The solid lines refer to the approximate bounding parabolas of ref. [1]. The insets show the coupling topologies of the Heisenberg spin systems. Spin centers (solid circles) connected by solid bonds are coupled with strength J, dashed bonds indicate a coupling with strength $J_1 = \alpha J$.

[1] Schmidt H.-J., Schnack J. and Luban M., *Europhys. Lett.*, **55** (2001) 105.
[2] Schnack J. and Luban M., *Phys. Rev. B*, **63** (2001) 014418.
[3] Ashcroft N. W. and Mermin N. D., *Solid State Physics* (Saunders College Publishing) 1976, chapter 33, equation (33.32).
[4] $\mathbf{S}_\mu^-$ is the lowering operator on site $\mu$. |F> denotes the state with $M = S_{max}$, the term "ferromagnetic state" the one with $S = S_{max}$, and "spin-wave states" those with $S = S_{max} - 1$.
[5] Waldmann O., *Phys. Rev. B*, **65** (2002) 024424.